\documentclass[12pt]{article}
\usepackage{graphicx}
\usepackage[utf8]{inputenc}
\usepackage[english
]{babel}
\def\baselinestretch{1.5}
\begin{document}
\begin{center} 
{\Large \bf{
Correlations in a system of classical--like coins simulating spin-1/2 states in the probability representation of quantum mechanics}}
\end{center}
\begin{center} 
{\bf V. N. Chernega$^1$, O. V. Man'ko$^{1,2}$, V. I. Man'ko$^{1,3}$}
\end{center}

\medskip

\begin{center}
$^1$ - {\it Lebedev Physical Institute, Russian Academy of Sciences\\
Leninskii Prospect 53, Moscow 119991, Russia}\\
$^2$ - {\it Bauman Moscow State Technical University\\
The 2nd Baumanskaya Str. 5, Moscow 105005, Russia}\\
$^3$ - {\it Moscow Institute of Physics and Technology (State University)\\
Institutskii per. 9, Dolgoprudnyi, Moscow Region 141700, Russia}\\
Corresponding author e-mail: omanko@sci.lebedev.ru
\end{center}

\begin{abstract}
An analog of classical ``hidden variables" for qubit states is presented. The states of qubit (two-level atom, spin-1/2 particle) are mapped onto the states of three classical--like coins. The bijective map of the states corresponds to the presence of correlations of random classical--like variables associated with the coin positions ``up'' or ``down'' and the observables are mapped onto quantum observables described by Hermitian matrices. The connection of the classical--coin statistics with the statistical properties of qubits is found.  
\end{abstract}

\section{Introduction}
In quantum mechanics, the formulation of quantum system states and quantum observables \cite{Dirac,Schrod,Landau,Neumann} uses the notion of state vectors $|\psi\rangle$ in the Hilbert space, the state density operators acting in the Hilbert space and quantum observables are identified with the Hermitian operators acting in this space. Different representations of the state vectors and density operators in the form of wave functions or density matrices as well as in the form of quasidistributions on the system phase space like Wigner function \cite{Wig32}, Husimi function \cite{Husimi40}, Glauber--Sudarshan function \cite{Glauber63,Sud563} were constructed. The probability representation of quantum states where the states are identified with fair probability distributions was introduced both for continious variables \cite{Mancini}-\cite{OVVILasRes} and discrete spin variables \cite{Dod,OlgaJETP,Bregence,Weigert1,Wiegert2,Painini}; see review \cite{MarmoPhysScr15t02015}. The problems of formulations of quantum mechanics in different representations are particularly associated with intension to find the formulation as close as possible to classical intuition and classical understanding what is the state and what is the observable in the classical physics.

The aim of this work is to present  the formulation of  notion of quantum states and observables on the example of spin--1/2 system (two--level atom, qubit), using the model of three classical--like coins and classical--like observables related to the games with the coins. In fact, we consider an analog of formal ``hidden variables'' for spin--1/2 system. The contemporary review of hidden variables in quantum mechanics is  given by Genovese \cite{Genovese}. We construct in explicit form the bijective map of the density matrix of qubit states and quantum observables described by Hermitian $2\times2$ matrices onto probability distributions describing the positions of the three coins ``up'' or ``down'' and the classical--like observables associated with the rules of the usual game with coin tossing, respectively. The geometry of the qubit state in this construction corresponds to the map of Bloch sphere geometry \cite{ChuangNelson} onto triangle geometry illustrated by the triada of Malevich's squares on the plane  \cite{Chernega1,Chernega2,Chernega3} and called quantum suprematism representation of spin--1/2 states \cite{PhysScr2018,Entr2018,confScr2018,MAVI2018Turin}. Different ideas to construct the formulation and geometry of quantum states closer to the classical picture of the system behavior were discussed earlier, e.g., in \cite{Wooters,Mielnik}.

This paper is organized as follows. In Section 2, we review the quantum suprematism approach to spin--1/2 states. In Section 3, we construct a map of the spin--1/2 (qubit, two--level atom) observable onto the classical--like coin observables. Also  we obtain the formulas connecting the quantum observable statistics with the classical--coin statistics in the presence of the coin probabilities and coin--observable  correlations. The conclusions are presented in Section 4.

\section{Qubit states in quantum suprematism picture}
The Hermitian density $2\times2$ matrix $\rho$ of spin--1/2 state in the basis $|m\rangle$, where $m=\pm1/2$ is the projection of spin onto the $z$ axis, reads
\begin{equation}\label{eq.1}
\rho=
\left(\begin{array}{cc}
\rho_{1/2, 1/2}&\rho_{1/2, -1/2}\\
\rho_{-1/2, 1/2}&\rho_{-1/2, -1/2}
\end{array}\right).
\end{equation}
Following \cite{Ventrig2017} and using the notation $p_1,\,p_2,\,p_3$ for probabilities to have in the state (\ref{eq.1}) the spin projections $m=+1/2$ onto axes $x,\,y,\,z,$ respectively, we can express these probabilities as
\begin{equation}\label{eq.2}
p_k=\mbox{Tr}\left(\rho\rho_k\right),\quad k=1,2,3,
\end{equation}
where $\rho_k=|\psi_k\rangle\langle \psi_k|$ are the density matrices of pure states with the state vectors 
\begin{equation}\label{eq.3}
|\psi_1\rangle=\left(\begin{array}{c}
1/\sqrt2\\1/\sqrt2\end{array}\right),\quad |\psi_2\rangle=\left(\begin{array}{c}
1/\sqrt2\\i/
\sqrt2\end{array}\right),\quad |\psi_3\rangle=\left(\begin{array}{c}
1\\0\end{array}\right).
\end{equation}
In view of (\ref{eq.2}) and (\ref{eq.3}), we obtain the expression for the density matrix (\ref{eq.1}) in terms of the probabilities $p_1,\,p_2,\,p_3$, i.e.,  
\begin{equation}\label{eq.4}
\rho=\left(\begin{array}{cc}
p_3&p_1-1/2-i(p_2-1/2)\\p_1-1/2+i(p_2-1/2)&1-p_3\end{array}\right).
\end{equation}
This relation means that we construct an invertable map of the density matrix $\rho$ onto the 3--vector with probability components $(p_1,p_2,p_3)=\vec{\cal P}$, i.e.,  
\begin{equation}\label{eq.5}
\rho\leftrightarrow\vec{\cal P}, 
\end{equation}
and the sum of the vector components must be equal to unity. This relation demonstrates that the spin--1/2 state is determined by three probability distributions given by the probability vectors
\begin{equation}\label{eq.6}
\vec P_1=(p_1,1-p_1),\,\vec P_2=(p_2,1-p_2),\,\vec P_3=(p_3,1-p_3).
\end{equation}
The probability vectors are not independent. Since the density matrix (\ref{eq.4}) must have nonnegative eigenvalues, the probabilities $p_1,\,p_2,\,p_3$ should satisfy the inequality 
\begin{equation}\label{eq.7}
(p_1-1/2)^2+(p_2-1/2)^2+(p_3-1/2)^2\leq 1/4.
\end{equation}
For pure states $\mbox{Tr}\rho^2\,=1$, the inequality is converted into the equality 
\begin{equation}\label{eq.8}
(p_1-1/2)^2+(p_2-1/2)^2=p_3(1-p_3).
\end{equation}
Since $p_3(1-p_3)=(p_3-1/2)^2-1/4$ the relation (\ref{eq.8}) is symmetric with respect to permutation $1\rightarrow2\rightarrow3$. Condition (\ref{eq.7}) reflects the presence of quantum correlations in the qubit system. 

Let us consider now three classical--like independent nonideal coins. Tossing these coins, we get three probability distributions (\ref{eq.6}). But for independent classical--like coins, the nonnegativity condition of matrix (\ref{eq.4}) does not valid. The quantumlike description of classical system states was studied in \cite{Koopman,Neumann1}; see also \cite{OVVILasRes}.

The states of classical systems can be associated with analogs of density operators \cite{OVVILasRes,Clmech} but there is no nonnegativity condition for these operators, and they can have negative eigenvalues. The probabilities satisfying inequality (\ref{eq.7}) and describing the spin--1/2 state belong to the ball and the surface of sphere of the radius 1/2  in the 3--dimensional space, with the center given by vector $\vec p_0$ with coordinates $\vec p_0=(1/2,1/2,1/2)$. The probabilities describing the possible states of three classical--like coins belong to the cube, i.e., $0\leq p_1\leq1,\,0\leq p_2\leq1$, $0\leq p_3\leq1$. There is no dependence of the different coin probabilities, i.e., there are no correlations providing inequality (\ref{eq.7}). But if one wants to simulate the quantum behavior of the spin--1/2 system, the corresponding correlations for the classical--like coins must be introduced. In this case, the map (\ref{eq.5}) of matrix (\ref{eq.4}) onto the vector $\vec{\cal P}$, where probabilities $p_1,\,p_2,\,p_3$ describe the classical--like coin states, provides the possibility of simulation of the spin--1/2 state behavior by the classical--like coins. Both the classical--like coin probability distributions (\ref{eq.6}) and probabilities determining quantum spin--1/2 state and satisfying inequality (\ref{eq.7}) are illustrated by the triadas of Malevich's squares \cite{Chernega1,Chernega2,Chernega3,PhysScr2018,Entr2018}. The squares have three sides determined by the probabilities $p_1,\,p_2,\,p_3$, and the length $y_k$ of the kth side reads 
\begin{equation}\label{eq.9}
y_k=[2p_k^2+2p_{k+1}+2p_k p_{k+1}-4p_k-2p_{k+1}+2]^{1/2}.
\end{equation}
The sum of the areas of three Malevich's squares is
\begin{eqnarray}
S(p_1,\,p_2,\,p_3)&&=2[3+2(p_1^2+p_2^2+p_3^2)-3(p_1+p_2+p_3)\nonumber\\
&&+p_1p_2+p_2p_3+p_3p_1].\label{eq.10}
\end{eqnarray}
For the classical--like coin states, the sum $S(p_1,\,p_2,\,p_3)$ has maximum value $S_{max}^{(c)}=6$. The maximum classical value of the sum (\ref{eq.10}) is reached for two cases where all probabilities are zero or one. For the spin--1/2 states, the maximum value of the area (\ref{eq.10}) is $S_{max}^{(1)}=3$ \cite{Entr2018}. This means that quantum correlations provide the constraints on the value of probabilities as well as the difference of the maximum value of the square area characteristics of the classical and quantum states. The picture of quantum probabilities in terms of quantum suprematism representation of the qubit states illustrates the difference of geometry of the classical--like coin states and spin--1/2 states. It is known that quantum correlations of two--qubit states provide the violation of Bell inequalities \cite{Bell64} characterized by the difference of maximum classical correlation parameters represented by the number 2 and quantum parameter given by the number $2\sqrt2$. We see that even in the qubit state the discussed quantum correlations provide the difference of independent classical--like coin system behavior and spin--1/2 system behavior simulated by the classical coins with extra constraints due to the difference of the square areas 6 and 3. 

The correlations can be detected in experiments with superconducting circuits based on Josephson junction devices \cite{ShuelKoin,Astafiev,Walraff} or in experiments with neutrons \cite{Venalabor}.  Corresponding measurements with the superconducting qubits, which are analogs of two--level atoms or spin--1/2 particles, also determine the maximum of the sum of areas of Malevich's squares. 

For this, one has to measure the spin projections, e.g., of neutron on three perpendicular directions. The obtained mean values of the spin projections $x_1,\,x_2,\,x_3$ determine the probabilities $p_1,\,p_2,\,p_3$, i.e., $p_k=(x_k+1)/2,\,k=1,\,2,\,3.$ The results of the measurement are used to find the maximum of sum (\ref{eq.10}) of Malevich's square area. This sum has to be compared with the theoretical number  3. 

\section{Quantum observables and classical--like variables}
In this section, we consider the simulation of quantum observables  for qubits by three dichotomic classical--like random variables. 

Let us define the rules of game with three classical--like coins as follows. If, due to tossing, the first coin has position ``up'' the gain equals $x$, for position ``down'' the loss is the same. Thus, the random variable 
$X=\left(\begin{array}{c}
x\\-x\end{array}\right),$ 
associated with the first coin, has two values, and for second coin the analogous random variable is 
$Y=\left(\begin{array}{c}
y\\-y\end{array}\right).$ 
For third coin we define gain and loss by the random variable 
$Z=\left(\begin{array}{c}
z_1\\z_2\end{array}\right)$ 
with two different values. The mean values of random variables are determined by the probability distributions $(p_1,1-p_1),\,(p_2,1-p_2),$ and $(p_3,1-p_3)$ as follows:  
\begin{eqnarray}
&&\langle X\rangle=p_1 x-x(1-p_1), \nonumber\\
&&\langle Y\rangle=p_2 y-y(1-p_2),\label{eq.11}\\
&&\langle Z\rangle=p_3 z_1+z_2(1-p_3).\nonumber
\end{eqnarray}
Let us rewrite the random variables $X,\,Y,\,Z$ in the form of $2\times2$-matrix 
\begin{equation}\label{eq.12}
A=\left(\begin{array}{cc}
z_1&x-iy\\
x+i y&z_2\end{array}\right).
\end{equation}
The matrix is an arbitrary Hermitian $2\times2$-matrix and can be used to simulate an arbitrary qubit observable. The qubit state has the  density matrix $\rho$ (\ref{eq.4}) expressed in terms of the probabilities $p_1,\,p_2,\,p_3$. The density matrix provides the possibility to calculate all the moments of arbitrary qubit observable (\ref{eq.12}), i.e.,  
\begin{equation}\label{eq.13}
\langle A^n\rangle=\mbox{Tr}(\rho A^n), \quad n=1,2,\,\ldots\,
\end{equation}
For example, as one can check the mean value of the observable $A$ has the form
\begin{equation}\label{eq.14}
\langle A\rangle=\langle X\rangle+\langle Y\rangle+\langle Z\rangle, 
\end{equation}
which is the sum of the mean values of introduced classical--like random variables. To obtain the highest moments $\langle A^n\rangle$, we use the generating function
\begin{equation}\label{eq.15}
G(\lambda)=\mbox{Tr}\left[\rho\exp\lambda A\right]=\sum_{n=0}^\infty\frac{\lambda^n}{n!}\langle A^n\rangle.
\end{equation}
Using the formula
\begin{equation}\label{eq.16}
\exp t(\vec\sigma\vec n)=(\cosh t)1_2+(\sinh t)(\vec\sigma \vec n), 
\end{equation}
where $1_2$ is the unity $2\times2$-matrix, $\sigma_x,\,\sigma_y,\,\sigma_z$ are Pauli matrices 
\begin{equation}\label{eq.17}
\sigma_x=\left(\begin{array}{cc}
0&1\\1&0\end{array}\right),\quad\sigma_y=\left(\begin{array}{cc}
0&-i\\i&0\end{array}\right),\quad\sigma_z=\left(\begin{array}{cc}
1&0\\0&-1\end{array}\right), 
\end{equation}
and $\vec n$ is the unit vector, i.e., $\vec n^2=1$, we obtain  
\begin{equation}\label{eq.18}
G(\lambda)=\exp\left(\lambda\frac{z_1+z_2}{2}\right)\mbox{Tr}\left[\rho\exp(\lambda r)(\vec\sigma\vec n)\right].
\end{equation}
Here $\vec n=\vec r/r,$ $ \vec r=(x,y,z),$ $z=(z_1-z_2)/2,$\\ $r=\sqrt{x^2+y^2+z^2}$ and 
\[\exp\lambda r(\vec \sigma\vec n )=\cosh\lambda r \left(\begin{array}{cc} 1&0\\0&1\end{array}\right)+(\sinh\lambda r)(\vec\sigma\vec n).\]
The statistics of quantum observable $A$ is determined by the highest moments
\begin{equation}\label{eq.16a}
\langle A^n\rangle=\frac{d^n G(\lambda)}{d\lambda^n}|_{\lambda=0}\,.
\end{equation}
Using formulas (\ref{eq.16a}), we obtain 
\begin{eqnarray}
&&\frac{d G(\lambda)}{d\lambda}=\frac{(z_1+z_2)}{2}G(\lambda)\nonumber\\
&&+r\exp\left(
\frac{\lambda(z_1+z_2)}{2}\right)\left[\sinh\lambda r+f\cosh\lambda r\right],\label{eq.17a}
\end{eqnarray}
where 
\begin{eqnarray}
&&f=r^{-1}\left[\langle A\rangle-\frac{z_1+z_2}{2}\right],\nonumber\\
&&\langle A\rangle=(2 p_1-1)x+(2p_2-1)y+p_3z_1+(1-p_3)z_2, \label{eq.18a}
\end{eqnarray}
and
\begin{equation}\label{eq.19a}
\frac{d^2G(\lambda)}{d\lambda^2}=(z_1+z_2)\frac{d G(\lambda)}{d\lambda}+\left[r^2-\left(\frac{z_1+z_2}{2}\right)^2\right]G(\lambda).
\end{equation}
One can check that
\begin{equation}\label{eq.20}
\langle A^2\rangle=(z_1+z_2)\langle A\rangle+\left[r^2-\left(\frac{z_1+z_2}{2}\right)^2\right].
\end{equation}
Due to (\ref{eq.19a}), all the derivatives of the generating function $
d^n G(\lambda)/d \lambda^n$ are expressed in terms of $G(\lambda)$ and $d G(\lambda)/d\lambda.$ 
We have the following property of the highest moments of quantum observable $A$. All the highest moments depend on the probabilities $p_1,\,p_2,\,p_3$ only due to the dependence of mean value $\langle A\rangle$ on these probabilities.

The obtained results can be formulated as the following recepie: How to simulate the quantum mechanics of spin--1/2 system by classical rules of game with three classical--like coins and classical--like variables $x,\,y,\,z_1,\,z_2$ associated with the coin tossing? One has probability vector $\vec{\cal P}=(p_1,p_2,p_3)$ as a result of tossing the coins. The vector is mapped onto the matrix $\rho$ which is postulated to be density matrix, and this means that there are quantum correlations expressed by inequality (\ref{eq.7}). Three classical random variables defined by the rules of the coin game and taking values $(x,-x);\,(y,-y);\,(z_1,z_2)$ are associated with the matrix (\ref{eq.12}). This matrix is postulated to be a qubit quantum observable. After this, applying the quantum rules of obtaining the statistics of quantum observable for given quantum states, we express all the highest moments of an arbitrary observable in terms of classical coin probabilities $p_1,p_2,p_3$ and classical random variables. Such quantum ingredient as the fidelity is expressed in terms of the probabilities associated with the classical coin game, i.e.,  
\begin{eqnarray}
&&\mbox{Tr}(\rho_1\rho_2)=p_3{\cal P}_3- (1-p_3)(1-{\cal P}_3)\nonumber\\
&&+\left[(p_1-1/2)-i(p_2-1/2)\right]\left[({\cal P}_1-1/2)+i({\cal P}_2-1/2)\right]\nonumber\\
&&+\left[(p_1-1/2)+i(p_2-1/2)\right]\left[({\cal P}_1-1/2)-i({\cal P}_2-1/2)\right].\nonumber\\
&&\label{eq.21}
\end{eqnarray}
Here  $p_1,p_2,p_3$ are the probabilities which determine the state with density matrix $\rho_1$ and ${\cal P}_1,\,{\cal P}_2,\,{\cal P}_3$ are the probabilities which determine the state $\rho_2$. Also such quantum property as the superposition principle can be formulated as nonlinear addition rule for probabilities determining the state, which are pure ones \cite{PhysScr2018,Cher22}. 

\section{Conclusions}
To conclude, we point out the main results of our work. We demonstrated that the quantum mechanics of such system as qubit (spin--1/2, two--level atom) can be simulated by using three classical--like coin  states associated with probabilities to get coin positions ``up'' and ``down''. Also quantum spin--1/2 observables can be simulated by the rules of game with these three coins. The quantumness of the system in this picture is related to the presence of quantum correlations imposed onto the coin behavior and expressed in terms of inequality (\ref{eq.7}). The state density matrices are constructed using the classical--like coin tossing probabilities by postulating the form of matrices (\ref{eq.4}). The new observation of this study is the existence of generating function (\ref{eq.15}) for highest moments of spin--1/2 observables and its expression in terms of classical--like coin probabilities and classical--like random variables. The approach has geometrical interpretation for the qubit states in terms of Malevich's square  picture. The quantumness of the states is responsible for the bound 3 for the maximal area of the sum of Malevich's squares. The developed method can be extended to the case of qudits. We present this method in the future publication. It is worth noting that the formalism of quantum mechanics and its relation to classical physics formalism is discussed in the literature during many decades. In this connection,  the review \cite{Mermin} presents the recent discussion of the approach called QBism \cite{Fuchs}. In addition to this, the quantum suprematism representation we used to discuss the example of spin--1/2 system states and observables in terms of absolutely classical objects like classical coins and classical random variables provides the possibility also to clarify some classical--quantum connections. In is worth noting that the probabilities $p_1,\,p_2,\,p_3$ do not satisfy the equation $p_1+p_2+p_3=1$. In fact, we use not joint probability distribution which gives the conditional probabilities with Bayes rule but the set of three probability distributions which obey the constraints (\ref{eq.7}).

\subsection*{Acknowledgments}
Vladimir I. Man'ko thanks Professor Tommaso Calarco for fruitful discussion of relations of Malevich's squares with qubit states.


\begin{thebibliography}{99}

\bibitem{Dirac}  P. Dirac, {\it The Principles of Quantum Mechanics} (Oxford University Press, 1930)
\bibitem{Schrod} E. Schr\"odinger, {\sl Ann. Phys. } (Leipzig), {\bf 79}, (1926) 489.
\bibitem{Landau}L. D. Landau, 
{\sl Z. Physik}, {\bf 45}, (1927) 430.
\bibitem{Neumann}J. von Neumann, 
{\sl Nach. Ges. Wiss.} G\"ottingen, {\bf 11}, (1927) 245.
\bibitem{Wig32} E. Wigner,  {\sl Phys. Rev.}, {\bf 40}, (1932) 749.
\bibitem{Husimi40} K. Husimi,  {\sl Proc. Phys. Math. Soc. Jpn.}, {\bf 23}, (1940) 264. 
\bibitem{Glauber63} R. J. Glauber, {\sl  Phys. Rev. Lett.}, {\bf10}, (1963) 84.
\bibitem{Sud563} E. C. G. Sudarshan,  {\sl Phys. Rev. Lett.}, {\bf 10}, (1963) 277. 
\bibitem{Mancini}S. Mancini, V. I. Man'ko,  P. Tombesi,  {\sl Phys. Lett. A}, {\bf 213}, (1996) 1.
\bibitem{Mancini1} S. Mancini, V. I. Man'ko,  P. Tombesi,  {\sl Found. Phys.}, {\bf 27}, (1997) 801.
\bibitem{OVVILasRes} O. V. Man'ko,  V. I. Man'ko,  {\sl J. Russ. Laser Res.}, {\bf 18}, (1997)  407.
\bibitem{Dod}V. V. Dodonov,  V. I. Man'ko,  {\sl Phys. Lett. A}, {\bf 229}, (1997) 335.
\bibitem{OlgaJETP}V. I. Man'ko,  O. V. Man'ko,  {\sl J. Exp. Theor. Phys.}, {\bf 85}, (1997) 430. 
\bibitem{Bregence}O. V. Man'ko, in: B.Gruber, and M. Ramek (Eds.), {\it Proceedings of
	International Conference ``Symmetries in Science X'' (Bregenz,
	Austria, 1997)} (Plenum Press, New York 1998) 207.
\bibitem{Weigert1}
S. Weigert,  {\bf Phys. Rev. Lett.}, {\bf 84}, (2000)  802.
\bibitem{Wiegert2} J.  P. Amiet,  S. Weigert,
{\sl J. Opt. B: Quantum Semiclass. Opt.}, {\bf1}, (1999)  L5.
\bibitem{Painini} G. M. D'Ariano, L. Maccone,  M. Paini, 
{\sl J. Opt. B: Quantum Semiclass. Opt.}, {\bf 5}, (2003) 77.
\bibitem{MarmoPhysScr15t02015}M. Asorey, A. Ibort, G. Marmo,  F. Ventriglia, {\sl  Phys. Scr.},  
{\bf90}, (2015) 074031.
\bibitem{Genovese} M. Genovese,  {\sl Phys. Rep.}, {\bf 413/6}, (2005) 319.
\bibitem{ChuangNelson}
M. A. Nielsen ,  I. L. Chuang, {\it Quantum Computation and Quantum Information}  (Cambridge University Press, Cambridge,
UK, 2000)
\bibitem{Chernega1} V. N. Chernega, O. V. Man'ko, V. I. Man'ko, {\sl J. Russ. Laser Res.},  {\bf 38}, (2017) 
141.
\bibitem{Chernega2} V. N. Chernega, O. V. Man'ko, V. I. Man'ko, {\sl J. Russ. Laser Res.},  {\bf38},  (2017) 
324.
\bibitem{Chernega3}  V.  N. Chernega, O. V. Man'ko, V. I. Man'ko, {\sl J. Russ. Laser Res.},  {\bf 38},  (2017) 416.
\bibitem{PhysScr2018} M. A. Man'ko, V. I. Man'ko,  {\sl Phys. Scr.}, {\bf 93}, (2018) 084002.
\bibitem{Entr2018}J. A. L\'opez-Saldivar, O. Casta\~nos, E. Nahmad-Achar,  R. L\'opez-Pe\~na, M. A. Man'ko, V. I. Man'ko,  {\sl Entropy}, {\bf20}, (2018) 630.
\bibitem{confScr2018} V. N. Chernega, O. V. Man'ko, V. I. Man'ko, {\sl J. Phys.: Conf. Ser.},  {\bf1071}, (2018) 012008.
\bibitem{MAVI2018Turin} M. A. Manko, V. I. Man'ko, {\sl J. Russ. Laser Res.}, {\bf39}, (2018) 1. 
\bibitem{Wooters} W. K. Wootters,  {\sl Found. Phys.}, {\bf16}, (1986) 391. 
\bibitem{Mielnik} B. Mielnik,  {\sl Commun. Math. Phys.}, { \bf 9}, (1968) 55.
\bibitem{Ventrig2017}V.  I. Man'ko, G. Marmo, F. Ventriglia, P. Vitale,  {\sl J. Phys. A: Math. Theor.}, {\bf500}, (2017) 335302.
\bibitem{Koopman} B. O. Koopman,  {\sl Proc. Natl. Acad. Sci. USA}, { \bf 17}, (1931) 315.
\bibitem{Neumann1} J. von Neumann,  {\sl Ann. Math.},{ \bf33}, (1932) 587.
\bibitem{Clmech}O. V. Man'ko, V. I. Man'ko,  {\sl J. Russ. Laser Res.}, { \bf25}, (2004) 477.
\bibitem{Bell64} J. S. Bell,  {\sl Physics}, {\bf 1}, (1964) 195.
\bibitem{ShuelKoin} M. H. Devoret, R. J. Schoelkopf, {\sl Nature}, { \bf406}, (2002) 1039. 
\bibitem{Astafiev} Yu. A. Pashkin, T. Yamamoto, O. Astafiev, Y. Nakamura, D. V. Averin, J. S. Tsai, {\sl Nature}, { \bf421}, (2003) 823.
\bibitem{Walraff} 
A. Wallraff ,  {\sl Phys. Rev. Lett.}, {\bf 106} (2011) 220503.
\bibitem{Venalabor} H. Rauch,  {\sl Contemporary Physics}, {\bf 27}, (1986) 345.
\bibitem{Cher22}V. N. Chernega, O. V. Man'ko, V. I. Man'ko,  {\sl J. Russ. Laser Res.},  {\bf 39}, (2018) 128.
\bibitem{Mermin}
N. David Mermin, {\it "Making Better Sense of Quantum Mechanics"}, arXiv:1809.01639 [quant-ph] (2018)
\bibitem{Fuchs}
Christopher A. Fuchs, Blake C. Stacey, {\it "QBism: Quantum Theory as a Hero's Handbook"},  arXiv:1612.07308 [quant-ph] (2017)


\end{thebibliography}
\end{document}